\documentclass[12pt]{article}
\usepackage{amsmath,amssymb,amscd,cite}
\newtheorem{thm}{Theorem}[section]

\newtheorem{lem}[thm]{Lemma}

\newtheorem{defi}[thm]{Definition}
\newcommand{\pf}{{\bf Proof. \ }}
\newcommand{\qed}{\hfill $\Box$ \\}

\font\msbm=msbm10 at 12pt

\newcommand{\F}{\mbox{\msbm F}}
\newtheorem{rem}[thm]{Remark}
\newtheorem{ex}[thm]{Example}

\begin{document}
\author{ Kenza Guenda and T. Aaron Gulliver
\thanks{T. Aaron Gulliver is with the
Department of Electrical and Computer Engineering,
University of Victoria, PO Box 3055, STN CSC, Victoria,
BC, Canada V8W 3P6. email: agullive@ece.uvic.ca.}}

\title{New Symmetric and Asymmetric Quantum Codes}
\date{}
\maketitle
\begin{abstract}
The asymmetric CSS construction is extended to the Hermitian case.
New infinite families of quantum symmetric and asymmetric codes are
constructed. The codes obtained are shown to have parameters
better than those of previous codes.
A number of known codes are special cases of the codes given here.
\end{abstract}

\section{Introduction}

Quantum codes have been introduced as an alternative to classical
codes for use in quantum communication channels. Since the landmark
result in \cite{shor} and \cite{steane96}, this field of research has grown rapidly.
In particular, classical codes have been used to construct good quantum codes~\cite{calderbank96}.
Until recently, most of the codes
have been designed under the assumption that the channel is
symmetric.
Ioffe and M\'ezard~\cite{ioffe} argued that in
physical systems the noise is typically asymmetric, i.e., qubit-flip
errors occur less frequently than phase-shift errors, so to design
efficient error correcting codes, this asymmetry should be exploited.
Using this fact, Ioffe and M\'ezard obtained asymmetric quantum error correcting codes from BCH and LDPC codes.
They showed that these codes have good performance.

The Calderbank-Shor-Steane (CSS) construction uses a pair of
classical codes, one for correcting qubit-flip errors and the other
for correcting phase-shift errors \cite{CS,S}.
The concept of asymmetric codes was introduced by Steane \cite{steane}.
Sarvepalli et al.~\cite{pradeep} extended the CSS construction to asymmetric
stabilizer codes. The classical codes were chosen such that the code
for phase errors has a larger distance than the code for qubit-flip
errors. Further, they showed the advantage of asymmetric quantum
codes over symmetric quantum codes, and gave several infinite
families of asymmetric quantum codes obtained from nested classical
codes. A mathematical framework for asymmetric quantum codes and
some construction results were given in~\cite{wang}. This framework
was extended to additive codes by Ezerman et
al.~\cite{ezerman1,ezerman2}.

In this paper, we extend the asymmetric CSS construction to the Hermitian case.
New infinite families of quantum codes are presented.
In addition, new symmetric and asymmetric quantum codes are obtained
from binary BCH codes.
These codes have known minimum distances, and the relationship
between the rate gain and minimum distance is given explicitly. The
minimum distances are larger than those of the codes given
in~\cite{sar}. The concatenation of MDS codes is used to obtain new
quantum codes. These codes have good parameters and provide
significantly more flexibility in code design than the construction
of Ezerman et al.~\cite{ezerman2}.

\section{Preliminaries}

For a positive integer $n$, let
$V_n=(\mathbb{C}^q)^{\otimes^n}=\mathbb{C}^{q^n}$, be the $n$th tensor product of $\mathbb{C}^q$.
Then $V_n$ has the following orthonormal basis $\{|c \rangle=|c_1,\ldots,c_n \rangle,
c=(c_1,\ldots,c_n) \in \F_q^n\}$. A $q-$ary quantum code of length
$n$ is a subspace $Q$ of $V_n$ of dimension $k$.
\begin{defi}
Let $d_x$ and $d_z$ be positive integers. A quantum code $Q$ in
$V_n$ with $K \geq 1$ codewords is called an asymmetric quantum code
(AQC) with parameters $((n,K,\{d_z,d_x\}))_q$ or
$[[n,k,\{d_z,d_x\}]]_q$ (with $k=\log_qK$), if $Q$ detects $d_x-1$
quantum symbols of $x$ errors and also $d_z-1$ quantum symbols of $z$ errors.
\end{defi}
\begin{lem}(Standard CSS Construction for AQCs)
\label{CSS}

(i) Let $\mathcal{C}_1$ and $\mathcal{C}_2 $ be two classical codes
over $\mathbb{F}_q$ with parameters $[n, k_1 , d_1 ]$ and $[n, k_2 ,
d_2]$, respectively, such that $C_1\subset C_2 \subset
\mathbb{F}_q$. Then there exists a quantum code with parameters
$[[n,k_2-k_1,\{d_z,d_x\}]]_q$, where $d_z = \max\{ wt(C_2 \setminus
C_1 ), wt (\mathcal{C}_1^{\bot} \setminus \mathcal{C}_2^{\bot})\}$
and $d_x = \min \{ wt(\mathcal{C}_2 \setminus \mathcal{C}_1), wt
(\mathcal{C}_1^{\bot} \setminus \mathcal{C}_2^{\bot})\}$.

(ii) If there exists two classical linear codes $\mathcal{C}_1$ and
$\mathcal{C}_2$ over $\mathbb{F}_{q^2}$ with parameters
$[n,k_1,d_1]_{q^2}$ and $[n,k_2,d_2]_{q^2}$ such that
$\mathcal{C}_1^{\bot h}\subset \mathcal{C}_2$, then there exists an
asymmetric quantum code with parameters $[[n, k_2-k_1^{h} ,
\{d_x,d_z\}]]_{q^2}$, where $\{d_x,d_z\}=\{d_1,d_2\}.$
\end{lem}
\pf Part (i) follows from the restatement of the CSS construction
for asymmetric stabilizer codes given in~\cite[Lemma 3.1]{pradeep}.
For (ii), we have that if a code is $\F_{q^2}$-linear then the
Hermitian dual is the same as the trace Hermitian dual~\cite[Theorem 3]{calderbank96}.
Hence from~\cite[Theorem 4.5]{ezerman1}, there exists an asymmetric code with parameters
$[[n,k_2-k_1^{h},\{d_z,d_x\}]]_{q^2}$, where $\{d_z,d_x\}=\{d_1,d_2\}$.
\qed

Note that an asymmetric quantum code with parameters
$((n,K,\{d,d\}))_q$ is a symmetric quantum code with parameters
$((n,K,d))_q$, but the converse is not true~\cite[Remark 2.3]{wang}.
An AQC is pure whenever $\{d_z,d_x\}=\{d_1,d_2\},$ otherwise it is
said to be degenerate.

A negacyclic code $C$ of length $n$ over
$\mathbb{F}_q$ such that $(n,q)=1$ is an ideal of the ring
$R=\mathbb{F}_q[x]/(x^n+1)$, generated by a polynomial $g(x)$ which divides $x^n+1$.
The code $C$ is uniquely determined by its defining
set $T=\{0 \leq i \leq 1 | \ g({\alpha}^i)=0\},$ where $\alpha$ is a $2n${th} primitive root of the unity.
The following lemma characterizes the defining set of these codes.
\begin{lem}
\label{lem:Th.4.4.9} (\cite[Lemma 13]{guenda011})
Let $O_n$ be the set of odd integers from
$1$ to $2n-1$. If $C$ is  negacyclic over $\mathbb{F}_{q^2}$, then
the Hermitian dual $C^{\bot h}$ is also negacyclic with defining
set $\{i \in O_n : i \notin -qT\}$.
\end{lem}

\begin{lem}
\label{lem:allone} (\cite[Theorem 6.1]{fred}) Let $C$ be a linear
$[n,k,d]_q$ code. If $C$ contains the all one codeword, then there
exists an asymmetric quantum code $Q$ with parameters
$[[n,k-1,\{d,2\}]]_q$.
\end{lem}
Using Lemma~\ref{lem:allone}, Chee et al.~\cite[Corollary 6.2]{fred}
constructed some asymmetric quantum maximum distance separable (AQMDS) codes.
This lemma can be used to obtain many AQCs
(not necessarily  MDS), as shown in the following.
\begin{thm}
\label{th:best} There exist asymmetric quantum codes with the
following parameters:
\begin{itemize}
\item[(i)] $[[n,k-1,\{d,2\}]]_q$, and $[[n-1,k-1,\{d-1,2\}]]_q$ with $n,k,d$ the parameters of a $q$-ary narrow-sense BCH code.
\item[(ii)] $[[n,n/2-1,\{d,2\}]]_2$, with $n,d$ the parameters of any binary self-dual code.
\item[(iii)] $[[2^m-1,m,\{2^{m-1}-1,2\}]]_2$.
\end{itemize}
\end{thm}
\pf For part (i), a narrow-sense classical BCH code of designed
distance $\delta$ over $\F_q$ is a cyclic code generated by
$g(x)=\mbox{lcm}(M_1,\ldots, M_{\delta-1})$ that contains the all
one codeword. Hence from Lemma~\ref{lem:allone}, there is a quantum
asymmetric code with parameters $[[n,k-1,\{d,2\}]]_q$. By puncturing
a narrow-sense BCH code, we obtain a linear code which also contains
the all one codeword. The dimension is $k$. Since a BCH code is
cyclic, its permutation group is transitive. Hence the minimum
distance has been decreased by one~\cite[Corollary 15 Ch.
8]{macwilliams}. For part (ii), it is well known that the dual of a
binary self-orthogonal code contains the all one codeword, hence the
result follows from Lemma~\ref{lem:allone}. For part (iii), the
simplex code $S_m$ is a code with parameters $[2^m-1,m,2^{m-1}]_2$.
It is a cyclic code generated by $x^n-1/M_1^*(x)$, where $M_1^*(x)$
is the reciprocal polynomial associated with $Cl(1)$, the cyclotomic
class of 1. This is a subcode of the cyclic code $\mathcal{C}_0$
generated by $f_1(x)=x^n-1/(x-1)M_1^*(x)$. $\mathcal{C}_0$ is a
$[2^m-1,m+1,2^{m-1}-1]$ code. The codeword $f_1(x)M_1^*(x)$ is equal
to the all one codeword and is in $\mathcal{C}_0$. Hence there
exists an asymmetric quantum code with parameters
$[[2^m-1,m,\{2^{m-1}-1,2\}]]_2$. \qed
\begin{ex}
Applying (i) in Theorem~\ref{th:best}, we obtain the
$[[15,k',\{d_z,2\}]]_4$ codes listed in Table 1 from the $[15,k, d]_4$
BCH codes. Note that these codes have parameters similar to those
in~\cite[Table 1]{ezerman2} obtained via the best known linear codes (BKLCs).
\begin{table}
\label{tab:self}
 \caption{$[[15,k',\{d_z,2\}]]_4$ codes obtained from BCH codes over $\mathbb{F}_4$.}
\begin{center}
{\small
\begin{tabular}{|c|c|c|c|c|c|c|}
\hline
 $k'$& 2&3&5&7&8&10\\
\hline
 $d_z$&11&10&7&6&5&3\\
 \hline
\end{tabular}}
\end{center}
\end{table}
\begin{table}
 \caption{$[[n,k,\{d_x,d_z\}]]_4$ Asymmetric quantum codes obtained from punctured BCH codes over $\mathbb{F}_4$.}
\begin{center}
{\small
\begin{tabular}{|c|c|c|}
\hline
 $[[14,6,\{6,2\}]]_4$&$[[20,9,\{6,2\}]]_4$& $[[32,8,\{10,2\}]]_4$\\
  $[[14,9,\{4,2\}]]_4$&$[[20,12,\{4,2\}]]_4$& $[[32,18,\{7,2\}]]_4$\\
 $[[30,21,\{4,2\}]]_4$&$[[30,16,\{6,2\}]]_4$ &$[[30,11,\{10,2\}]]_4$\\
 $[[34,8,\{6,2\}]]_4$&$[[34,17,\{4,2\}]]_4$& ]$[[34,23,\{2,2\}]]_4$\\
 $[[38,27,\{2,2\}]]_4$&$[[38,21,\{8,2\}]]_4$& $[[38,15,\{9,2\}]]_4$\\
 $[[38,9,\{12,2\}]]_4$&$[[40,11,\{19,2\}]]_4$& $[[40,21,\{8,2\}]]_4$\\
$[[44,31,\{4,2\}]]_4$&$[[44,26,\{6,2\}]]_4$& $[[44,20,\{8,2\}]]_4$\\
$[[44,15,\{10,2\}]]_4$&$[[44,9,\{12,2\}]]_4$& $[[50,27,\{8,2\}]]_4$\\
$[[50,23,\{13,2\}]]_4$&$[[50,19,\{16,2\}]]_4$& $[[62,39,\{10,2\}]]_4$\\
$[[62,27,\{20,2\}]]_4$&$[[62,11,\{30,2\}]]_4$& $[[62,8,\{41,2\}]]_4$\\
$[[64,9,\{38,2\}]]_4$&$[[64,11,\{12,2\}]]_4$ &$[[64,17,\{12,2\}]]_4$ \\
$[[64,29,\{12,2\}]]_4$&$[[64,35,\{10,2\}]]_4$&$[[64,47,\{5,2\}]]_4$ \\
 \hline
\end{tabular}}
\end{center}
\end{table}
For example, we have the existence of codes with parameters $[[64,29,\{12,2\}]]_4$ and $[[64,9,\{38,2\}]]_4.$
Note that in~\cite[Table 2]{ezerman2}, codes with parameters
$[[64,29,\{\ge 4,2\}]]_4$ and $[[64,9,\{\ge 24,2\}]]_4$ were obtained using the
concatenated RS code construction.
\end{ex}

\subsection{Quantum BCH Codes}

The idea of constructing CSS codes from BCH codes is not new, as
Steane~\cite{steane99} and Aly et al.~\cite{aly07,sar} give quantum
symmetric and asymmetric codes. The difficulty with their methods is
the lack of knowledge of the dual distances, as noted
in~\cite{pradeep}. In addition, there is a rate gain which is
linearly proportional to the reduction in the distance of the code
used for correcting qubit-flip errors \cite[Lemma 4.6]{pradeep}.
Here we give quantum BCH codes with known minimum distances that
have a rate gain larger than the codes in \cite[Lemma 4.6]{pradeep}.
\begin{lem}
\label{lem:BCH1} Let $n=2^{m}-1$ and $B(\delta_i)$ be the BCH code
of designed distance $\delta_i$ such that $2 \le \delta_1 \le
\delta_2 \le 2^{\lceil \frac{m}{2} \rceil}-1 $. Then there exists an
AQC with the following parameters
\[ [[n,n+m-m\left(\frac{\delta_1+\delta_2}{2}\right),\{d_z,d_x\}]]_2,\]
with $d_z=wt(B(\delta_2))$ and $d_x=wt(B(\delta_1))$.
\end{lem}
\pf The condition $\delta_i \le 2^{\lceil \frac{m}{2}\rceil}-1$
gives that the dimension of $B(\delta)$ is equal to
$2^m-1-m\frac{\delta-1}{2}$ from~\cite[Corollary
9.3.8]{macwilliams}. We also have from~\cite[Lemma 1]{steane99} that
$B(\delta)^{\bot} \subset B(\delta)$. The minimum distance of
$B(\delta_i)^{\bot}$ is larger than
$2^{m-1}-2^{\frac{m}{2}}(\frac{\delta_i-1}{2})$ by the
Carlitz-Uchiyama bound~\cite[Theorem 9.9.18]{macwilliams}, and the
Singleton bound gives that $wt(B(\delta_i)) \le m
(\frac{\delta_i-1}{2})+1$. Hence the result follows from
Lemma~\ref{CSS} by taking $C_1=B^{\bot}(\delta_2)$ and
$C_2=B(\delta_1)$. \qed

In Table 3, we compare the codes obtained using
Lemma~\ref{lem:BCH1} and the construction in~\cite[Table 2]{sar},
when using the same BCH codes.
\begin{table}
\label{tab:com}
 \caption{Quantum Code Comparison}
\begin{center}
{\small
\begin{tabular}{|c|c|}
\hline
New asymmetric codes& Asymmetric codes in~\cite{sar}\\
\hline
$[[1023,803,\{31,15\}]]_2$ &$[[1023,80,\{32,15\}]]_2$\\
$[[1023,823,\{31,11\}]]_2$&$[[1023,100,\{32,11\}]]_2$\\
$[[1023,843,\{31,7\}]]_2$&$[[1023,120,\{32,7\}]]_2$\\
$[[1023,863,\{31,3\}]]_2$&$[[1023,140,\{32,3\}]]_2$\\
\hline
\end{tabular}}
\end{center}
\end{table}
\begin{rem}
By Lemma~\ref{lem:BCH1}, we have the existence of AQCs with parameters
$[[511,304,\{31,17\}]]_2$ and $[[255,183,\{15,5\}]]_2$.
From~\cite[Table 2, Table 3]{sar}, we have the existence of AQCs with parameters
$[[1023,70,\{32,17\}]]_2$ and $[[255,159,\{17,5\}]]_2$ from BCH codes and LDPC codes.
Table 3 shows that the construction provides codes with good rate gain.
\end{rem}
\begin{lem}
Let $n=2^m-1$, $m$ odd, such that $gcd(i,m)=1$, where $ 2^{i}+1 \le
2^{\lceil \frac{m}{2} \rceil} -1$. Furthermore, let $B_i$ be the
cyclic code with defining set $Cl(1)\cup Cl(2^i+1)$, and
$B(\delta_i)$ the BCH code with designed distance $\delta_i=2^i+1$.
Then there exist AQCs with parameters
\begin{equation}
\label{eq:charpin} [[2^m-1,2^{m}-1-m(2+2^{i-1}),\{d_{z_1},5\}]]_2,
[[2^m-1,m(2^{i-1}-2),\{d_{z_2},5\}]]_2,
\end{equation}
where $d_{z_1}=wt(B(\delta_i))$ and
$d_{z_2}=wt(B(\delta_i)^{\bot}\setminus B_i^{\bot})$.

When $q$ is a prime power, there exists AQCs with the
following parameters
\[
[[2q-1,2(k_1-k_2),\{q-k_1,k_2+1\}]]_q.
\]
\end{lem}
\pf The binary cyclic code $B_i$ with defining set $ Cl(1)\cup
Cl(2^i+1)\}$ (related to the Preparata codes), has parameters
$[2^m-1,2^m-2m-1,5]$, and
$wt({B_i}^{\bot})=2^{m-1}-2^{\frac{m-1}{2}}$ from ~\cite[Theorem
4.15]{charpin}. Consider the binary BCH code $B(\delta _i)$ with
$\delta_i=2^i+1 \le 2^{\lceil \frac{m}{2} \rceil}-1$. Hence the
dimension of $B(\delta)$ is equal to $2^m-1-m2^{i-1}$. By the BCH
bound, the minimum distance of $B(\delta_i)$ is at least $2^i+1$. The
Carlitz-Uchiyama bound gives that $wt(B_i\setminus
B(\delta_i)^{\bot})=5$. By the Singleton bound we have
$wt(B(\delta_i)) \le m(2^{i-1})+1$, so that $wt(B(\delta_i)\setminus
B_i^{\bot})= wt(B(\delta_i)$. Then~(\ref{eq:charpin}) follows from
Lemma~\ref{CSS} by taking $C_2=B_i$ and $C_1=B(\delta_i)^{\bot}$ in
the first case, and $C_2=B_i$ and $C_1=B(\delta_i)$ in the second
case.

For the last part, consider two RS codes $C_1$ with parameters
$[q-1,k_1,q-k_1]_q$ and $C_2$ with parameters $[q-1,k_2,q-k_2]_q$,
such that $k_1>k_2$. Extending $C_i$ we obtain $\overline{C}_i$.
Taking the direct sum $\mathcal{C}_i=C_i \bigoplus \overline{C}_i$
results in codes with parameters $[2q-1,2k_i,q-k_i]_q$.
From the generator and parity check matrices, we can easily verify that
${\mathcal{C}_i}^{\bot}={C_i}^{\bot} \bigoplus
{\overline{C}_i}^{\bot}$. Hence by Lemma~\ref{CSS} we obtain an AQC
with parameters $[[2q-1,2(k_1-k_2),q-k_1/k_2+1]]_q$. \qed

\begin{ex}
There exist AQCs from RS codes with the following parameters
\[
[[31,14,\{7,3\}]]_{16}, [[31,4,\{14,2\}]]_{16} \mbox{ and }[[31,22,\{4,3\}]]_{16}.
\]
\end{ex}
\subsection{Concatenated Construction}

In~\cite{ezerman2}, Ezerman et al. used the concatenated
construction~\cite[Ch 10]{macwilliams} for RS codes to obtain AQCs
over $\F_4$ with $d_x=2$, so the emphasis was only on $d_z$. Thus
their construction is limited in terms of the possible code
parameters. Therefore, we first provide a modification of their
construction.

Let $q$ be a prime power and $\F_{q^m}$ a finite extension of
$\F_{q}$. The trace of an element $\alpha$ in $\F_{q^m}$ is defined
as $Tr_{\F_{q^m}
/\F_{q}}(\alpha)=Tr_m(\alpha)=\sum_{i=0}^{m-1}\alpha ^{q^i}$.

Two bases $B=\{\alpha_1,\alpha_2,\ldots,\alpha_m \}$ and $B'=\{\beta_1,\beta_2,\ldots,\beta_m \}$ of $\F_{q^m}$ over
$\F_{q}$ are called dual bases if they satisfy
\[
Tr_{m}(\alpha_i \beta_j)=\left\{
\begin{array}{ll}
 1& \text{ if }i=j, \\
 0 & \text{ otherwise}. \\
\end{array}
\right.
\]
If a basis is the dual of itself, it is called a self-dual basis.

Let $B=\{\alpha_1,\alpha_2,\ldots,\alpha_m\}$ be a basis of $\F_{q^m}$
over $\F_{q}$. Then any $x \in \F_{q^m}$ can be written uniquely as
$x = \sum_{i=1}^{m}a_i\alpha_{i},$ with $a_i \in \F_{q}$. Define the
mapping
\begin{eqnarray}
\label{hom}\Psi_B : \F_{q^m} &\to& \F_{q}^m \\
x&\mapsto& \Psi_B(x)=(a_1,\ldots,a_m).
\end{eqnarray}
This is an $\F_q$-linear mapping which is bijective and can be
extended to
\begin{eqnarray}
\label{homo}\Phi_B : \F_{q^m}^n &\to& \F_{q}^{mn} \\
(x_1,x_2,\ldots,x_n)&\mapsto& (\Psi_B(x_1),\ldots,\Psi_B(x_m)).
\end{eqnarray}
\begin{lem}
\label{lem:rs} Let $B$ be a basis of $\F_{q^m}$ over $\F_q$ and $C$
an $[n,k,d]$ code over $\F_{q^m}$. Hence the code $\Phi_B(C)=\mathcal{C}$
is an $[nm,km, D\ge d]_q$ code. If $C$ is an $[n,k,n-k+1]_{q^m}$ MDS code,
then there exists a code $\widetilde{\mathcal{C}}$ over $\F_q$ with
parameters
\[[(m+1)n,km,2(n-k+1)]_q.\]
\end{lem}
\pf The first result is obvious, thus we prove only the second result.
For this, let $C$ be an $[n,k,n-k+1]_{q^m}$ MDS code. Then $\Phi_B(C)$ is
a code with parameters $[mn,mk,\ge d]_{q}$. This can be improved by
adding an overall parity check to each element $\Psi_B(x)$
to obtain a $q$-ary code $\widetilde{\mathcal{C}}$ with parameters
$[(m+1)n,km,2(n-k+1)]_q$~\cite[Ch. 18.8]{macwilliams}. \qed
\begin{lem}
\label{lem:self} Let $C$ be a linear code of length $n$ over
$\F_{q^m}$, and $\Phi$ the mapping defined in (\ref{homo}).
In addition, let $B=\{\alpha_1, \ldots, \alpha_m \}$ a basis of $\F_{q^m}$ over
$\F_{q}$, and $B^{\bot}=\{\beta_1,\beta_2,\ldots,\beta_m \}$ its
dual basis. Then we have
$\Phi_{B^{\bot}}(C^{\bot})=\Phi_B(C)^{\bot}$ and
$\Phi_{B^{\bot}}(C^{\bot h})=\Phi_B(C)^{\bot h}$.
\end{lem}
\pf Assume that $c'=(y_1,\ldots,y_n)\in C^{\bot}$. Hence $c\cdot
c'=0$ for all $c=(x_1,\ldots,x_n) \in C$, where $x_i=\sum_{j=1}^m
a_{ij}\alpha_j$ and $y_i=\sum_{j=1}^m b_{ij}\beta_j$. Then $c\cdot
c'=\sum_{i=1}^{n}(\sum_{j=1}^m a_{ij}\alpha_j)( \sum_{j=1}^m
b_{ij}\alpha_j)=0$.
The fact that $c\cdot c'=0$ implies that $Tr_{m}(c\cdot c')=0$, therefore
\[
Tr_{m}\left(\sum_{i=1}^{n}\left(\sum_{j=1}^m a_{ij}\alpha_j\right)\left( \sum_{j=1}^m b_{ij}\beta_j\right)\right)=0.
\]
Applying the additivity property of the trace
map, we obtain that this expression is equal to
\[
\sum_{i=1}^{n}Tr_{m}\left(\left(\sum_{j=1}^m a_{ij}\alpha_j\right)\left( \sum_{j=1}^m b_{ij}\beta_j\right)\right)=0.
\]
From the additivity of the trace map and the duality of the basis, we have that
\begin{equation}
\label{eq:du} Tr_m(c\cdot c')= \sum_{i=1}^{n}\sum_{j=1}^m
a_{ij}b_{ij}=0.
\end{equation}
Since
\[
\Phi_B(c)=(\Psi_B(x_1),\ldots,\Psi_B(x_n))=(a_{11},\ldots
a_{1m},a_{21},\ldots,a_{2m}, \ldots, a_{n1}, \ldots, a_{nm}),
\]
and
\[
\Phi_{B^{\bot}}(c')=(\Psi_{B^{\bot}}(y_1),\ldots,\Psi_{B^{\bot}}(y_n))=(b_{11},\ldots
b_{1m},b_{21}, \ldots ,b_{2m}, \ldots, b_{n1}, \ldots, b_{nm}),
\]
from (\ref{eq:du}) we obtain
\[
\Phi_B(c)\cdot\Phi_{B^{\bot}}(c')=\sum_{i=1}^{n}\sum_{j=1}^m
a_{ij}b_{ij}=0.
\]
Then $\Phi_{B^{\bot}}(C^{\bot}) \subset \Phi_B(C)^{\bot}$, and since
both codes have the same dimension the result follows.
The above proof also holds in the Hermitian case.
\qed

It is well known that the dual of an MDS code is also MDS. Thus if
$C_k$ is the RS code $[q^m-1,k,n-k]_{q^m}$, its dual $C_k^{\bot}$ is
a $[q^m-1,q^m-k,k+1]_{q^m}$ code. Consider now the code
$\Phi_{B^{\bot}}(C_k^{\bot})$ which from Lemma~\ref{lem:self} is
equal to $\Phi_B(C_k)^{\bot}$. This is a code with parameters
$[m(q^m-1),m(q^m-k), \ge k+1]_q$. Applying the construction of
Lemma~\ref{lem:rs}, we obtain that $\mathcal{C}^{\bot}_k$ has
parameters $[(m+1)(q^m-1),m(q^m-k), 2(k+1)]_q$. Since $\Phi$ is a
bijective map, if $k_1>k_2$ then $\Phi_B(C_{k_2}) \subset
\Phi_B(C_{k_1})$, and hence $\widetilde{\mathcal{C}}_{k_2} \subset
\widetilde{\mathcal{C}}_{k_1}$. Applying Lemma~\ref{CSS}, we obtain
the following result.
\begin{thm}
\label{th:conca} Let $q$ be an even prime power or $q$ an odd prime
and $m$ odd. Then there exist AQCs with parameters
$[[(m+1)(q^m-1),m(k_1-k_2),\{2(q^m-k_1),2(k_2+1)\}]]_q$.
\end{thm}

\noindent
In Table 4, we give some codes constructed from Theorem~\ref{th:conca}.
Note that the codes of length 45 have better parameters than the codes given by
Ezerman et al.~\cite[Table IX]{ezerman1}.
\begin{table}
\label{tab:co2m}
 \caption{Asymmetric quantum codes constructed from RS codes by Theorem~\ref{th:conca}}
\begin{center}
{\small
\begin{tabular}{|c|c|c||}
\hline
 $[[45,24,\{6,4\}]]_4$ & $[[45,24,\{8,2\}]]_4$ &$[[45,22,\{8,4\}]]_4$\\
 $[[45,16,\{14,4\}]]_4$ & $[[45,10,\{20,4\}]]_4$& $[[45,10,\{16,8\}]]_4$\\
 \hline
 $[[186,150,\{4,2\}]]_2$&$[[186,110,\{12,10\}]]_2$&$[[186,100,\{18,6\}]]_2$\\
 $[[186,80,\{24,10\}]]_2$&$[[186,45,\{34,16\}]]_2$&$[[186,40,\{44,6\}]]_2$\\
\hline
\end{tabular}}
\end{center}
\end{table}

In a similar way as for classical codes, a concatenated quantum
code~\cite{grassl} is constructed using two quantum codes, an outer
code $C$ and an inner code $C'.$ If $C$ is an $((n,K,d)))_q$ code,
then $C'$ must be an $((n',K',d'))_K$ code, i.e., $C'$ is a subspace
of $\mathbb{C}^{K^{\otimes^{n'}}}$. The concatenated code
$\mathcal{Q}$ is constructed as follows. For any codeword $|c>=
\sum_{i_1,\ldots, i_{n'}} \alpha_{i_1,\ldots,
i_{n'}}|{i_1,\ldots,i_{n'}}>$ in $C$, replace each basis vector
$|ij>$, where $ij=0,\ldots, K-1$ for $j=1, \ldots n'$, with a basis
vector $|x_{ij}>$ in $C'$, i.e., $|c> \mapsto |c">=
\sum_{i_1,\ldots, i_{n'}} \alpha_{i_1,\ldots,i_{n'}}|x_{i_1},\ldots,
x_{i_{n'}}>$. Hence the resulting code $\mathcal{Q}$ is an
$((nn',K', D))_q$ code with $D \ge dd'$.
\begin{thm}
Let $m$ be an integer, $q$ a prime power, and $m$, $k_1$, $k_2$ and $k$
positive integers less than or equal to $q^m-3$.
Then there exists a quantum code with parameters $[[(m+1)(q^m-1)((q^m-2),m(k_1-k_2),\ge
D]]_q.$ $D=dd'$ where $d'=\min\{2(q^m-k_1),2(k_2+1)\}$,
$d=\min\{2(q^m-k-1),k\}$.
\end{thm}
\pf From~\cite[Proposition 4.2]{fred}, we have the existence of
AQMDS $[[q^m-2,1,\{q^m-k-1,k\}]]_q$ codes for $k \le q^m-3$.
Consider the concatenated code $\mathcal{Q}$ with inner code the AQMDS
$[[q^m-2,1,\{q^m-k-1,k\}]]_q$ code. The outer code is the asymmetric
quantum code given in Theorem~\ref{th:conca}. The above
concatenation of quantum codes gives the result. \qed

In the remainder of this section, we use the previous
construction in the Hermitian case. We start by proving the
following lemma.
Assume that $q=p^2$ and that $q$ is even or $q$ and $m$ are odd.
\begin{lem}
\label{th:2011} Let $q=p^2$ be an odd prime power such that $q
\equiv 1 \mod 4$, $n$ an even divisor of $q-1$, and $s\le n-1$ an
even integer. Then there exists an MDS negacyclic code $C_s$ over
$\F_{q^2}$ of length $n$ with defining set
\[
T_s= \{i \text{ odd}: 1 \leq i \leq s-1 \},
\]
parameters $[n,n-\frac{s}{2},\frac{s}{2}+1]_{q^2}$ and which satisfies
$C^{\bot h}_s \subset C_s$.
\end{lem}
\pf  When $s= n-1$, from~\cite[Theorem 16]{guenda011} we obtain that $C_n$
is a negacyclic MDS code which is Hermitian self-dual.
Assume now that $s \le n-1$ is an even integer and $C_s$ is the negacyclic code with
defining set $ T_s= \{i \text{ odd}: 1 \leq i \leq s-1 \}$.
Hence the BCH bound~\cite{nuh}
gives that $C_s$
has minimum distance $\frac{s}{2}+1$. The dimension of $C_s$ is $n-|T_s|=n-\frac{s}{2}$.
This gives that $C_s$ is an $[n,n-\frac{s}{2},\frac{s}{2}+1]_{q^2}$ code.
The codes $C$ and $C_s$ are such that $C \subset C_s$, and hence $C_s^{\bot h} \subset {C_n}^{\bot h}$.
The Hermitian dual $C_s^{\bot h}$ is a negacyclic MDS code with
parameters $[n,\frac{s}{2},n-\frac{s}{2}+1]_{q^2}$. Since $C_n$ is
Hermitian self-dual, $C_s^{\bot h}\subset C_n \subset C_s$.
\qed

Assume now that we have a self-dual basis. Seroussi and
Lempel~\cite{lempel} proved that $\F_{q^m}$ has a self-dual basis
over $\F_{q}$ if and only if $q$ is even or both $q$ and $m$ are
odd. Consider the code $\Phi_{B^{\bot}}(C_s^{\bot h})$ which from
Lemma~\ref{lem:self} is equal to $\Phi_B(C_s)^{\bot h}$.
This is a code with parameters $[mn,m(n-\frac{s}{2}), \ge n- \frac{s}{2}+1]$.
Applying the construction of Lemma~\ref{lem:rs}, we obtain that
$\mathcal{C}^{\bot h}_s$ has parameters $[(m+1)n,m(n-\frac{s}{2}), 2(n-\frac{s}{2}+1)]_q$.
Since $\Phi$ is a bijective map and $C_s^{\bot h} \subset C_s$, then
$\Phi_B(C_s^{\bot h}) \subset \Phi_B(C_s^{\bot h})$, and hence
$\widetilde{\mathcal{C}}_s^{\bot h} \subset \widetilde{\mathcal{C}}_s$.
Applying Lemma~\ref{CSS}, we obtain the following result.
\begin{thm}
\label{lem:self1} Let $q=p^2$ be an even prime power or $q$ an odd
prime and $m$ odd. Then there exist AQCs with parameters
$[(m+1)n,m(n-s),\{2(\frac{s}{2}+1),2(n-\frac{s}{2}+1)\}]]_q$.
\end{thm}

\end{document}